\documentstyle[12pt]{article}
\begin{document}
\def\theequation{\arabic{section}.\arabic{equation}}
\newcommand{\be}{\begin{equation}}
\newcommand{\ee}{\end{equation}}
\begin{titlepage}
\title{Nonstationary Gravitational Lenses and the Fermat Principle}
\author{Valerio Faraoni \\ 
\\{\small \it S.I.S.S.A./I.S.A.S., Strada Costiera 11, 34014 Trieste
(Italy)}\\
{\small Present address: \it Department of Physics and Astronomy,
University of Victoria}\\
{\small \it P.O. Box 3055, Victoria, B.C. V8W~3P6 (Canada)}}
\date{
\maketitle
\thispagestyle{empty}
\begin{abstract}

We apply Perlick's (1990a) rigorous formulation of the Fermat principle in
arbitrary spacetimes to prove the correctness of the description
of gravitational lensing by gravitational waves, given in the
literature using the scalar and vector formalisms. We obtain
an expression for the time delay due to such nonstationary lenses; the
advantage over previous papers is that Perlick's formulation of the Fermat
principle is very rigorous and more suitable for practical calculations
in some cases. It is also shown that ordinary moving gravitational lenses must
be
considered as a stationary case.

\end{abstract}
\vspace*{0.5truecm}
\begin{center}
to be published in {\em The Astrophysical Journal}
\end{center}
\end{titlepage}
\clearpage
\setcounter{page}{1}
\section{Introduction}

The theory of gravitational lensing (e.g. Blandford \& Kochanek 1988,
and references therein) has been developed for {\em stationary}
gravitational lenses, mainly using the so-called vector and scalar
formalisms. The vector formalism (Bourassa, Kantowski \& Norton 1973;
Bourassa \& Kantowski 1975)
requires explicitly that the Newtonian potential $ \Phi $
describing the gravitational lens is static, $
\partial \Phi / \partial t =0 $. On the other hand, the
basis of the scalar formalism (Schneider 1985; Blandford \& Narayan
1986) is the Fermat principle which, in the
formulation employed (Weyl 1917; Quan 1962; Brill 1973), is
valid only for stationary spacetimes. This in turn implies the condition
$ \partial \Phi/\partial t =0 $ for the
Newtonian potential describing the lens, since
the post-Newtonian metric constructed out of $ \Phi $ is adopted as
the background. However, also {\em nonstationary}
lenses have been considered in the literature: cosmic strings (Hogan \&
Narayan 1984; Vilenkin 1986; Paczynski 1986) and gravitational waves,
both considered as lens components superimposed to the caustic
structure of an ordinary gravitational lens (McBreen \& Metcalfe 1988;
Allen 1989, 1990; Kovner 1990), or
as full lenses (Wheeler 1960; Zipoy 1966; Winterberg 1968; Zipoy
\& Bertotti 1968; Bertotti 1971, 1987; Kaufmann 1970; Bergmann 1971;
Bertotti \& Catenacci 1975; Burke 1975; Dautcourt 1975; Linder 1986,
1988; Braginsky et al. 1990; Faraoni 1992). In particular, Mc Breen
\& Metcalfe
have proposed that $\gamma$-ray bursts could result from the
amplification of small hot cores of Bl Lac objects by ordinary
microlensing, the microcaustics being jiggled by gravitational waves,
a suggestion criticized by Kovner (1990). In this paper we analize the
validity of the use of the vector and scalar formalisms in the
description of nonstationary
gravitational lenses, in particular gravitational waves.
Though the so-called propagation formalism (e.g. Blandford \& Kochanek
1988) based
on the optical scalar equations (Sachs 1961; Penrose 1966) has been
used in the past to describe propagation of light
through gravitational waves
(Bertotti 1971; Bertotti \& Catenacci 1975),
we will not take it into account, since it is not adequate to describe
image configuration, time delay and other features of gravitational
lenses that are important in order to compare theory with the observations.

As far as the scalar formalism is concerned, the version of the Fermat
principle employed in gravitational lens theory has been
generalized to nonstationary spacetimes by Kovner
(1990). A subsequent version by Perlick (1990a) does not change
Kovner's result, but strengthens his derivation of the principle by
using a more rigorous treatment, and provides explicit formulations more
suitable for practical calculations when isolated pulses of
gravitational radiation are considered (actually Kovner's calculations
have been performed for a stochastic background of gravitational
waves and, due to averaging, his formulae cannot be compared directly
with ours). Perlick's version is quite general; it requires neither
stationarity, nor causality restrictions on the spacetime structure.
Perlick has applied it to ordinary gravitational lenses and
to the conformally stationary case (1990b), but not to the (fully
nonstationary) situation of lensing gravitational waves. In
\S~2 we will make
use of Perlick's refined formulation to study the applicability of
the vector and scalar formalisms to the description of
lensing by gravitational waves, both from the theoretical point of
view, in order to improve and complete previous studies on this subject,
and in view
of future astrophysical applications. We will find that the scalar and
vector formalisms are still valid, provided one substitutes the usual
expressions for the arrival time functional and the deflection angle
with appropriate ones.

Before doing this, we will consider the (much more common) case of
moving ordinary gravitational lenses (stars or galaxies). The proper
motion of stars, or the rotation of a double star gravitational lens
can be responsible for microlensing events, and
similar motions can affect macrolensing by galaxies, though with
different effects. These phenomena are commonly described using the
vector or scalar formalism (Watson 1989; Schneider 1990; Kayser 1992; and
references therein). To the author's
knowledge, the validity of the hypothesis of lens stationarity needed
to apply these formalisms has never been tested in the literature, for such
cases, assuming it to hold {\em a priori}. However, it is easy to
realize that proper motions of stars or
galaxies, or rotation of a binary lens composed of such objects, are
not fast enough to make the lens nonstationary, and the condition $
\partial \Phi / \partial t $ holds with a very good
approximation. Though it is immediate to reach such a conclusion, we
will comment a little about it, because of its importance\footnote{We will
use units in which $ G=c=1 $ (but we will restore $ G $ and $ c $ in the
final formulae); the metric signature is +2; Greek indices run from 0 to 3,
and Latin indices run from 1 to 3.}.

We can test the validity of the stationary lens hypothesis by
comparing the crossing time of the lens $ \tau_c $ by a photon, with
the time scale of variation of the lens $ \tau_v $ (cf. Kovner 1990). The
ratio $ \tau_c/\tau_v $ measures the nonstationarity of the lens.

The value of $ \tau_c $ can be
estimated as the time required for a photon to cross an Einstein ring
centered on the lens, with radius $ r_e\equiv \sqrt{4MD} $,
where $ D \equiv
D_L D_{LS}/D_{S} $, and $ D_L $, $D_{LS} $ and $ D_S $ are the
observer-lens, lens-source and observer-source distances respectively.
The Einstein ring would be the image of a point source, if the lensing
mass were concentrated in a single point located exactly in front of
it. The Einstein ring radius can be regarded as the effective size of
a compact object lens; thus\setcounter{equation}{0}
\be
\label{1}
\tau_c \simeq \sqrt{4MD} \;.
\ee
The time scale of variation of the lens is of order $ \tau_v \sim
r_e/v $, where $ v $ is the (linear) velocity of the lens. We have
\be
\label{2}
\frac{\tau_c}{\tau_v} \simeq \frac{v}{c} <<1
\ee
for proper motions of stars in galaxies (typical velocities $ v\sim
100$ Km/s), or of galaxies in clusters ($ v\sim 1000$~Km/s).
Thus, we can neglect the time derivative of the potential describing
the Newtonian field of such lenses with a very good approximation.

Let us consider now a binary gravitational lens (Schneider \& Weiss
1986; Nemiroff 1988; Kochanek \& Apostolakis
1988; Erdl 1989; Mao \& Paczynski 1991). For such lenses
$ \tau_c \simeq r_e $, while $ \tau_v $ is of the order of the
orbital period $ P $ of
the binary system. Consider first an extragalactic binary system composed
of two common stars with masses $ M_1\sim M_2 \sim M_\odot $ and $
D \sim 300 $~Mpc; then $ \tau_c \sim 10 $~days. For a galactic such binary
system, with $ D \sim 1 $~Kpc, one has $ \tau_c \sim 20 $~minutes. These
values of $ \tau_c $ have to be compared with typical values of $
\tau_v\sim P $ ranging from months to years for the most common stars,
giving again $ \tau_c / \tau_v <<1$, and then $ \partial \Phi /\partial t
\simeq 0 $. Binary systems composed of massive objects with very short
orbital periods would have higher values of the ratio $ \tau_c/\tau_v
$; consider a binary sistem of black holes with $ M_1 \sim M_2 $ and
separation $ r $; from the third Kepler's law we have
\be
\frac{\tau_c}{\tau_v}\sim \frac{r_e}{P} \sim 0.2 \; \left(
\frac{M}{M_{\odot}} \right) \, \left( \frac{D}{{\mbox 1~Kpc}}
\right)^{1/2} \, \left( \frac{r}{R_{\odot}} \right)^{-3/2} \; .
\ee
Taking $ M \sim $100~$M_{\odot} $, $ D\sim $100~Kpc and $ r \sim
$10~$ R_{\odot} $, we get $ \tau_c/\tau_v \sim 1 $. Thus, it is
conceivable to construct a nonstationary lens with ``ordinary''
objects; however, such systems will be extremely rare in comparison with
the typical stars considered in studies on microlensing, and probably
are statistically insignificant.

Let us consider now a binary system of two similar galaxies in a cluster;
we assume that the mass of each galaxy contained within a radius $ R $
is given by (Persic \& Salucci 1992)
\be
M(R)=10^{13} \,\, M_{\odot} \,\left( \frac{R}{{\mbox 1 Mpc}}
\right)^{1.2} \, h^{-1} \; ,
\label{4}
\ee
where $ h $ is the Hubble constant in units of $ 50 $~$\mbox{km}$~${\mbox
s}^{-1}\,\mbox{Mpc}^{-1}$. This
gives (taking the radius $ R $ roughly equal to the separation $ r $
of the two galaxies)
\be
\frac{\tau_c}{\tau_v} \simeq 2 \cdot 10^{-6} \left( \frac{D}{\mbox{
100 Mpc}} \right)^{1/2} \, \left( \frac{r}{\mbox{1 Mpc}}
\right)^{-0.3}\,h^{-1} \; ,
\ee
so that we can again regard the assumption $ \partial \Phi /\partial t $
as satisfied with a very good degree of accuracy.

In the following, we will turn our attention to lensing gravitational
waves, which are a real case of nonstationary lenses.

\section{Lensing by gravitational waves and the Fermat principle}

For a photon deflected by a gravitational wave, the crossing time of
the lens is $ \tau_{c} \sim P $, the period of the gravitational wave,
while the variation scale of the lensing field is $ \tau_v \sim P $.
While for ordinary gravitational lenses $ \tau_v >> \tau_{c} $, in
this case $ \tau_v \sim \tau_{c} $, and the lens is nonstationary. In what
follows, we will restrict ourselves to consider localized pulses of
gravitational radiation, and will not consider the gravitational wave
background.

Let us consider the scalar formalism first; we assume that:
\begin{itemize}
\item the lens is geometrically thin (or, in other words, that
$\tau_c \sim P $ is much smaller than the Hubble time)
\footnote{For the most likely realistic situations, the region in which the
wave amplitudes differ appreciably from zero has size much smaller than the
path travelled by the photons (which can be a cosmological length), since
these amplitudes decrease with
the distance from the source of gravitational radiation. Moreover, the
thin lens approximation is correct for waves which propagate along the
direction orthogonal to the source-observer direction; troubles could
arise as the angle $ \theta $ between the propagation direction of the wave
and the observer-source line increases, the situation getting worst at
$ \theta =\pi /2 $. Note however that, even for this extremal
case, either the light
never reaches the gravitational wave and there is no lensing at all,
or the light and the gravitational wave move the one towards the
other. In this last case, photons are not deflected, as one infers
from the papers by Zipoy (1966) and Linder (1986), or from our
eq.~(\ref{deflectionangle}): for a wave that can be locally
approximated by a plane wave propagating along the $z$-axis (i.e.
parallely to the photon's path), it is immediate to see in the
transverse-traceless gauge that the deflection angle vanishes.};
\item the spacetime metric is given by\setcounter{equation}{0}
\be
\label{5}
g_{\mu\nu}=\eta_{\mu\nu}+h_{\mu\nu}
\ee
in an asymptotically Cartesian coordinate system $ \{x^{\mu}\} $, where
$ \eta_{\mu\nu} $ are the components of the Minkowski metric, and
$ h_{\mu\nu} $ are small perturbations;
\item geometric optics holds;
\item deflection angles are small.
\end{itemize}
Under these assumptions, the actual light
rays differ from straight lines only within the gravitational field of
the wave, i.e. in a region much smaller than the distance travelled by
the photons. This permits us to substitute the true photon path in the
3-dimensional space with a zig-zag path composed of two straight lines
from the source to the lens, and from the lens to the observer. This
zig-zag construction is employed by Blandford \& Kochanek (1988) and by
Kovner (1990); we will discuss later the validity of this substitution.

The zig-zag path has tangent vector
\be
\dot{x}^{\mu}( \lambda )=\left\{
\begin{array}{ll}
\left( \dot{x}^{0}( \lambda ), \dot{\xi}( \lambda ) \, \vec{n}_{ol} \right) &
\mbox{ if $\lambda< \lambda_{l} $}  \nonumber \\
 \left( \dot{x}^{0}( \lambda ), \dot{\xi}( \lambda ) \, \vec{n}_{ls}
\right) & \mbox{ if $ \lambda>\lambda_{l} $} \;,\nonumber
\end{array}  \right.
\label{6}
\ee
where $ \lambda $ is a parameter along the photon path (with value $
\lambda_{l} $ at the lens position), and $ \vec{n}_{ol} $ and $
\vec{n}_{ls} $ are three-dimensional vectors satisfying
\be
\label{7}
{\vec{n}_{ol}}^{2}={\vec{n}_{ls}}^{2}=1 \; ,
\ee
pointing in the lens-observer and in the source-lens directions
respectively. The zig-zag path is assumed to be a null curve
satisfying the on shell condition
\be
\label{onshell}
g_{\mu \nu}\, \dot{x}^{\mu}\dot{x}^{\nu}=0 \;,
\ee
which gives a second degree equation for the variable
$ \dot{\xi}/\dot{x}^{0} $,
\be
\left( \frac{\dot{\xi}}{\dot{x}^{0}}\right)^{2}\left( 1+h_{ij}n^{i}n^{j}
\right) +2\, h_{0i}n^{i} \left( \frac{\dot{\xi}}{\dot{x}^{0}}\right) +
 h_{00}-1 =0 \; ,
\label{9}
\ee
whose solutions are
\be
\label{10}
\frac{\dot{\xi}}{\dot{x}^{0}}=\frac{
-h_{0i}n^{i}\pm
\sqrt{(h_{0i}n^{i})^{2}+(1-h_{00})(1+h_{ij}n^{i}n^{j})}}{1+h_{ij}n^{i}n^{j}}
\;.
\ee
Choosing the ``$+$'' sign (corresponding to $ d\xi/dx^{0} >0 $, i.e.
to photons travelling in the direction of increasing $ \xi $), we get
\be
\label{11}
\frac{\dot{\xi}}{\dot{x}^{0}}=1-\frac{1}{2} \left(
h_{00}+2h_{0i}n^{i}+h_{ij}n^{i}n^{j} \right) +O(h^{2}) \; .
\ee
We set up the geometry as customary in gravitational lens theory; we
consider a source and a lens concentrated at single redshifts in the
source plane and in the lens plane, with coordinates
$ \underline{s}=(s_x,s_y) $ and $ \underline{x}=(x,y) $, respectively.
Though we are
considering a Minkowskian background, a cosmological model can be
taken into account simply by allowing the $D$'s to be angular diameter
distances in a Friedmann-Lemaitre-Robertson-Walker
universe\footnote{In addition, we have to multiply the final time
delay in eq.~(\ref{ttilde}) by $ 1+z_L $, where $ z_L $ is the
redshift of the lens plane, to get the time delay at the observer's
position.}.

Perlick's (1990a) first coordinate version of the Fermat principle is

{\bf Theorem}: let $ (M, g_{ab})$ be a spacetime; let $ U \subseteq M $ be
an open set on
which a local ($\cal{C}^{\infty}$) coordinate system
$ \{ x^0,x^1,x^2,x^3 \} $ is defined, such that $ \partial_{0} $ is
timelike; a future pointing null $\cal{C}^{\infty}$ curve $ \sigma :[a,b]
\rightarrow U $ with coordinate representation $ \lambda \mapsto x^{i}(
\lambda) $ is a geodesic iff it makes the functional\footnote{The integral
is computed along the curve $ \sigma $ from the source to the
observer. The differential form of eq.~(\ref{functional})
coincides with the corresponding eq.~(84.5) of Landau \& Lifshitz (1975).}
\be
\label{functional}
\omega( \sigma ) \equiv \int_{S}^{O} \left[ \left( \hat{g}_{ij}
\dot{x}^{i}\dot{x}^{j} \right) ^{1/2} -\hat{\phi}_{i}\dot{x}^{i}
\right] ( \lambda ) \, d\lambda
\ee
stationary with respect to
($\cal{C}^{\infty}$) variations satisfying
\begin{eqnarray}
\label{13}
\delta x^{\mu}( S)=0  \\
\delta x^{i}( O)=0
\label{14}
\end{eqnarray}
\be
\label{15}
 \delta \left( \dot{x}^{0}-\left( \hat{g}_{ij}\dot{x}^{i}\dot{x}^{j}
\right) ^{1/2}+ {\hat{\phi}}_{i}\dot{x}^{i} \right) =0 \; ,
\ee
where
\begin{eqnarray}
\label{16}
f & \equiv & \frac{1}{2}\, \ln |g_{00}| \; , \\
\label{17}
\hat{\phi}_{i} & \equiv & -e^{-2f} g_{0i} \; , \\
\hat{g}_{ij} & \equiv & e^{-2f} g_{ij}+\hat{\phi}_{i}\hat{\phi}_{j}  \; .
\label{18}
\end{eqnarray}
The null zig-zag paths fail to be $\cal{C}^{\infty} $ curves at the lens
position $ \lambda_l $; however, this does not affect the
validity of the theorem. They satisfy, to first order in the
gravitational wave amplitudes $ h $, the conditions required to the varied
curves in the theorem:
\be
\label{19}
  \delta x^{\mu}( S)=0 \; ,
\ee
\be
\label{20}
  \delta x^{i}( O)=0  \; ,
\ee
and, using eqs.~(\ref{16})-(\ref{18}) and (\ref{onshell}):
\begin{eqnarray}
& & \delta \left( \dot{x}^{0}-\left( \hat{g}_{ij}\dot{x}^{i}\dot{x}^{j}
\right) ^{1/2}+ \hat{\phi}_{i}\dot{x}^{i} \right)=
\delta \left( \dot{x}^{0}-
\dot{\xi} \sqrt{ \left(
\frac{h_{0i}n^{i}}{1-h_{00}}\right)^{2}+\frac{1+h_{ij}n^{i}n^{j}}{1-h_{00}}
} -\dot{\xi} \, \frac{h_{0i}n^{i}}{1-h_{00}} \right) = \nonumber \\
& & =0+O(h^{2}) \; .
\label{21}
\end{eqnarray}
This proves that substituting the real null geodesics with our null
zig-zag paths is correct to first order in $ h $, provided the new
paths extremize the functional eq.~(\ref{functional}).

Computing this functional for our zig-zag paths, we get
\begin{eqnarray}
\omega & = & \int_{S}^{O} \dot{\xi} \left[  \sqrt{
\left( \frac{h_{0i}n^{i}}{1-h_{00}}\right)^{2}+\frac{1+h_{ij}n^{i}n^{j}}
{1-h_{00}}
}+\frac{h_{0i}n^{i}}{1-h_{00}} \right] ( \lambda) \, d\lambda = \nonumber \\
& = & \int_{S}^{O} \dot{\xi} \left[ 1+\frac{1}{2} \left(
h_{00}+2\, h_{0i}n^{i}+h_{ij}n^{i}n^{j} \right) \right] ( \lambda) \;
d\lambda +O(h^{2})= \nonumber \\
& = & \int_{S}^{O} \dot{x}^{0} \, d\lambda +O(h^2) \; ,
\label{22}
\end{eqnarray}
where eq.~(\ref{onshell}) has been used.

The zig-zag paths approximating the real null geodesics are extrema of
the travel time
\begin{eqnarray}
 & & \int_{S}^{O} dx^{0}=
\int_{S}^{O} \frac{\dot{x}^{0}}{\dot{\xi}} \,\, d\xi=  \nonumber \\
&  &= \int_{S}^{O} \left[ 1+\frac{1}{2}(
h_{00}+2h_{0i}n^{i}+h_{ij}n^{i}n^{j}) \right] d\xi +O(h^{2}) \; .
\label{23}
\end{eqnarray}
We can subtract from this functional the travel time calculated along a
straight line were the lens absent; in fact this quantity is a
constant (with value $ D_S $), and its variation vanishes identically. The
result is easily found to be
\be
\tilde{t}= \frac{1}{2cD}\left(
\underline{x}-\underline{s}\right)^{2}+\frac{{\cal H}}{c}
+O(h^{2}) \;,
\label{ttilde}
\ee
where
\be
{\cal H} =\frac{1}{2}\int_{S}^{O}
( h_{00}+2h_{0i}n^{i}+h_{ij}n^{i}n^{j} ) \, d\xi \;.
\label{25}
\ee
The functional $ \tilde{t} $ results from two contributions (compare
with the case of ordinary gravitational lenses, e.g. in Blandford
\& Kochanek (1988)); the first term in eq.~(\ref{ttilde}) is called
the {\em geometrical time delay}, and results from the extra length
travelled by the photon due to the deflection, when compared to the
propagation of light in the absence of the lens. The second term in
eq.~(\ref{ttilde}) is the {\em gravitational time delay} resulting from
the wave's field; its analog for the case of a Schwarzschild lens
is familiar from experiments in the Solar System (Will 1981, and
references therein).

We require stationarity of the ``arrival time functional'' $
\tilde{t} $ seen as a function of the lens plane coordinates,
\be
\label{26}
{\underline{\nabla}}_{\, x}\tilde{t}=0 \;,
\ee
and get the lens equation
\be
\label{27}
\underline{s}=\underline{\nabla}_{\,x} T( \underline{x}, t) \; ,
\ee
where
\be
\label{28}
T( \underline{x}, t) \equiv \frac{{x}^{2}}{2}+D\,{\cal H} (
\underline{x},t) \; .
\ee
If we assume that the unperturbed photon path lies along the $ z $-axis, we
can substitute the integral along the zig-zag path in
eq.~(\ref{25}) with the integral along the $ z $-axis, getting
\be
\label{29}
{\cal H}=\frac{1}{2}\int_{z_S}^{z_O} dz \, (h_{00}+2h_{03}+h_{33})
+O(h^2) \;;
\ee
the lens equation~(\ref{27}) can now be written (to first order)
as
\be
\underline{s}=\underline{x}+\frac{1}{2}\,\, {\underline{\nabla}}_{\,x}
\int_{S}^{O}dz \,  ( h_{00}+2h_{03}+h_{33}) .
\label{30}
\ee
Thus, we find that lensing by gravitational waves can be described by the
scalar formalism based on the appropriate version of the Fermat principle,
getting a lens equation in which the deflection angle is given by the
integral term in eq.~(\ref{30}).

\section{Discussion and conclusion}
 \setcounter{equation}{0}

Ordinary moving gravitational lenses must be considered as stationary,
as is usually done in the literature. On the contrary, gravitational
waves are essentially nonstationary lenses, and the validity of the
usual vector and scalar formalisms must be examined in such
situations. However, as seen in the previous section, applying the scalar
formalism to describe lensing by gravitational waves is correct to first
order, provided that the expression for the time delay
caused by the lens  is given by the functional
eq.~(\ref{ttilde}). This conclusion is not completely new, since it
obviously underlies Kovner's paper, but our work is based on Perlick's
formulation of the Fermat principle, which is easier to handle for practical
purposes for isolated pulses of gravitational waves, and is derived more
rigorously from the mathematical point of view. Moreover, the Fermat
principle makes us able to write the explicit expression of the
gravitational time delay caused by such a lens, and to prove the validity
of approximating the real null geodesics with zig-zag paths in the
nonstationary case.

As far as the vector formalism is concerned, it follows immediately
from the results of the previous section that the description of the
lens action as a mapping from the lens to the source plane used in the
vector formalism, and made explicit by the lens
equation\setcounter{equation}{0}
\be
\label{31}
\underline{s}=\underline{x}-D\underline{\alpha}
\ee
is correct also in our nonstationary case, provided one assumes, for a
photon whose unperturbed path is
along the $ z $-axis, the following expression for the deflection angle
\be
\underline{\alpha}=- \,\frac{1}{2}\, \underline{\nabla}_{\,x} \int_S^O dz\,
\left( h_{00}+2h_{03}+h_{33} \right) +O(h^2) \; .
\label{deflectionangle}
\ee
We must note that eq.~(\ref{deflectionangle}) can be obtained independently
from our considerations on the Fermat principle, simply by considering the
equation of null  geodesics for photons which suffer small deflections
from the $ z $-axis, due to the presence of gravitational waves
localized in a region of dimensions much smaller than the length
travelled by the photon. Equation~(\ref{deflectionangle}) has been derived
in this way by Linder (1986, 1988) and Bertotti (1987). This is consistent
with the observation that the hypothesis of lens
stationarity required in the vector formalism (Bourassa,
Kantowski \& Norton 1973), as it is usually
formulated for ordinary (stationary) gravitational lenses, is needed
only to compute the deflection angle, and thus one might think that
providing the correct expression of the deflection in the nonstationary
case will allow us to still use the formalism. However, this appears
to be  an heuristic argument supporting our conclusion, rather than a
real proof. Moreover, the use of the Fermat
principle and the scalar formalism provides us a deeper physical
insight, and a more elegant and rigorous formulation of the theory.

\section*{Acknowledgments}

It is a pleasure to thank Professor B. Bertotti for having pointed out
the idea of lensing by gravitational waves and for helpful
discussions, Drs. M. Persic and P. Salucci for stimulating
conversations, and an anonymous referee for useful remarks. This work
was supported by the Italian Ministero dell' Universit\`a e della
Ricerca Scientifica e Tecnologica.

\bigskip
\section*{References}

Allen, B. 1989, Phys. Rev. Lett. 63, 2017 \\
---------- 1990, Gen. Rel. Grav. 22, 1447 \\
Bergmann, P. G. 1971, Phys. Rev. Lett. 26, 1398 \\
Bertotti, B. 1971, in {\em General Relativity and Cosmology}, ed. B. K.
Sachs (New York: Academic Press), p.~347 \\
----------------  1987, private communication \\
Bertotti, B. \& Catenacci, R. 1975, Gen. Rel. Grav. 6, 329 \\
Blandford, R. D. \& Kochanek, C. S. 1988, in {\em
Proceedings of the 13th Jerusalem Winter School on Dark Matter in the
Universe}, ed. J. N. Bachall, T. Piran, \& S. Weinberg (Singapore: World
Scientific), p.~133 \\
Blandford, R. D. \& Narayan, R. 1986, ApJ 310, 568 \\
Bourassa, R. R., Kantowski, R. \& Norton, T. D. 1973, ApJ 185, 747 \\
Bourassa, R. R. \& Kantowski, R. 1975, ApJ 195, 13 \\
Braginsky, V. B., Kardashev, N. S., Polnarev, A. G. \& Novikov, I. D.
1990, Nuovo Cimento 105~B, 1141 \\
Brill, D. R. 1973, in {\em Relativity, Cosmology and Astrophysics},
ed. W. Israel (Dordrecht: Reidel), p.~134 \\
Burke, W. L. 1975, ApJ 196, 329 \\
Dautcourt, G. 1974, in {\em Confrontation of Cosmological
Theories with Observation}, IAU Symposium No. 63, ed. M. S. Longair
(Dordrecht: Reidel), p.~299 \\
Erdl, H. 1989, in {\em Cosmology and Gravitational Lensing},
Proceedings, Ringberg Castle, Germany 1989, ed. G. B\"{o}rner, T.
Buchert \& P. Schneider, MPA/P3 preprint, p.~78 \\
Faraoni, V. 1992, in {\em Gravitational Lenses}, Proceedings, Hamburg,
Germany 1991, ed. R. Kayser, T. Schramm \& S. Refsdal (Berlin: Springer
Verlag), in preparation \\
Hogan, C. \& Narayan, R. 1984, MNRAS 211, 575 \\
Kaufmann, W. J. 1970, Nature 227, 157 \\
Kayser, R. 1992, in {\em Gravitational Lenses}, Proceedings, Hamburg,
Germany 1991, ed. R. Kayser, T. Schramm \& S. Refsdal (Berlin:
Springer Verlag), in preparation \\
Kochanek, K. S. \& Apostolakis, J. 1988, MNRAS 235, 1073 \\
Kovner, I. 1990, ApJ 351, 114 \\
Landau, L. D. \& Lifshitz, E. M. 1975, {\em The Classical Theory of
Fields}, fourth revised edition (Oxford: Pergamon Press) \\
Linder, E. V. 1986, Phys. Rev. D~34, 1759 \\
---------------- 1988, ApJ 328, 77 \\
Mao, S. \& Paczynski, B. 1991, Princeton Observatory preprint
POP-399 \\
Mc Breen, B. \& Metcalfe, L. 1988, Nature 332, 234 \\
Nemiroff, R. J. 1988, Ap\&SS 145, 53 \\
Paczynski, B. 1986, Nature 319, 567 \\
Penrose, R. 1966, in {\em Perspectives in Geometry and Relativity,
Essays in Honor of V\`{a}clav Hlavat\'{y}},
ed. B. Hoffmann (Bloomington: University of Indiana Press), p.~259.\\
Perlick, V. 1990a, Class. Quantum Grav. 7, 1319 \\
------------- 1990b, Class. Quantum Grav. 7, 1849 \\
Persic, M. \& Salucci, P. 1992, in {\em The Distribution of Matter in
the Universe}, Proceedings of the second DAEC meeting, Meudon 1991, in
press \\
Quan, P. M. 1962, in {\em Les Th\'{e}ories Relativistes de la
Gravitation}, Proc. Royamount Conf. 1959 (Paris: CNRS), p.~165 \\
Sachs, R. 1961,  Proc. Roy. Soc. Lon. 264, 309 \\
Schneider, P. 1985, A\&A 143, 413 \\
---------------- 1990, in {\em Gravitational Lensing}, Proceedings,
Toulouse, France 1989, ed. Y. Mellier, B. Fort \& G. Soucail
(Berlin: Springer Verlag), p.~175 \\
Schneider, P. \& Weiss, A. 1986, A\&A 164, 237 \\
Vilenkin, A. 1984, ApJ (Letters) 282, L51 \\
Watson, W. D. 1989, in {\em Gravitational Lenses}, Proceedings,
Cambridge, USA 1988, ed. J. M. Moran, J. N. Hewitt \& K. Y. Lo
(Berlin: Springer Verlag), p.~195 \\
Weyl, H. 1917, Ann. Physik 54, 117 \\
Wheeler, J. A. 1960, in {\em Rendiconti della Scuola Internazionale di
Fisica ``Enrico Fermi''}, 11th Course of the Varenna Summer School, 1959,
``Interazioni Deboli'', ed. L. A. Radicati (Bologna: Zanichelli),
p.~67 \\
Will, C. M. 1981, {\em Theory and Experiment in Gravitational
Physics} (New York: Wiley) \\
Winterberg, F. 1968, Nuovo Cimento 53~B, 195 \\
Zipoy, D. M. 1966, Phys. Rev. 142, 825 \\
Zipoy, D. M. \& Bertotti, B. 1968, Nuovo Cimento 56~B, 195
\end{document}